\documentclass[11pt,psamsfonts,reqno,a4paper]{amsart} 
\usepackage[utf8]{inputenc}
\usepackage{mathtools,amsmath,amssymb}
\usepackage{graphicx}
\usepackage{lmodern}
\usepackage[T1]{fontenc}
\usepackage{microtype}
\usepackage{hyperref}
\usepackage{xspace}
\usepackage{bibentry}
\usepackage{easybmat}
\usepackage{mathdots}

\usepackage{cite}
\usepackage{graphicx}

\usepackage{subcaption}

\usepackage{bibentry}
\usepackage[margin=2.5cm]{geometry}

\usepackage[usenames,dvipsnames,svgnames,table]{xcolor}
\definecolor{todocolor}{HTML}{D7E1E5}
\usepackage[textsize=tiny, 
            backgroundcolor=todocolor, 
            bordercolor=todocolor, 
            linecolor=todocolor, 
            textwidth=3.9cm]{todonotes}

\usepackage{comment}
\usepackage{rotating}
\usepackage{tikz}
\usetikzlibrary{decorations.pathreplacing}
\numberwithin{equation}{section}
\newcommand{\osc}[1]{\mathbf{#1}}

\newcommand{\oa}{\osc{a}}

\newcommand{\oad}{\osc{\bar{a}}}

\newcommand{\NN}{\mathbf{N}}

\newcommand{\Lspi}{\mathbf{L}}

\newcommand{\vp}{\bar{\mathbf{v}}}
\newcommand{\vm}{\mathbf{v}}

\renewcommand{\wp}{\bar{\mathbf{w}}}
\newcommand{\wm}{\mathbf{w}}

\newcommand{\ap}{\bar{\mathbf{A}}}
\newcommand{\am}{{\mathbf{A}}}

\newcommand{\idb}{\JD }
\newcommand{\id}{{\ID }}

\newcommand{\ka}{\kappa}

\newcommand{\R}{R}
\newcommand{\RR}{\mathbf{R}}
\renewcommand{\L}{\mathrm{L}}

\DeclareMathOperator{\diag}{diag}

\newmuskip\pFqmuskip
\newcommand*\pFq[6][8]{%
    \begingroup
    \pFqmuskip=#1mu\relax
    \mathcode`\,=\string"8000
    \begingroup\lccode`\~=`\,
    \lowercase{\endgroup\let~}\pFqcomma
    {}_{#2}F_{#3}{\left(\genfrac..{0pt}{}{#4}{#5};#6\right)}%
    \endgroup}
\newcommand{\pFqcomma}{\mskip\pFqmuskip}

\def\be{\begin{eqnarray}}
\def\ee{\end{eqnarray}}

\DeclareMathOperator{\tr}{tr}

\newcommand{\Kop}{\mathrm{K}}
\newcommand{\Qop}{\mathrm{Q}}
\newcommand{\Pop}{\mathrm{P}}
\newcommand{\ID}{\mathrm{I}}
\newcommand{\JD}{\mathrm{J}}
\newcommand{\Sop}{\mathrm{S}}
\newcommand{\Bop}{\mathrm{B}}
\newcommand{\Gop}{\mathrm{G}}
\newcommand{\eL}{L^{\mathbf{6}}}

\makeatletter
\g@addto@macro\bfseries{\boldmath}
\makeatother

\makeatletter
\newcases{rrcases}{\quad}{%
  \hfil$\m@th\displaystyle{##}$}{$\m@th\displaystyle{##}$\hfil}{\lbrace}{.}
\makeatother

\begin{document}

\begin{abstract}
\noindent
% We initiate the study of operatorial Q-systems 
We present a family of novel Lax operators corresponding to representations of the RTT-realisation of the Yangian associated with $D$-type Lie algebras. These  Lax operators are of oscillator type, i.e. one space of the operators is infinite-dimensional while the other is in the first fundamental representation of $\mathfrak{so}(2r)$. We use the isomorphism between the first fundamental representation of $D_3$ and the $\mathbf{6}$ of $A_3$, for which the degenerate oscillator type Lax matrices are known, to derive the Lax operators for $r=3$. The results are used to generalise the Lax matrices to arbitrary rank for representations corresponding to the extremal nodes of the simply laced   Dynkin diagram of $D_r$. The multiplicity of independent solutions at each extremal node is given by the dimension of the fundamental representation. We further derive certain factorisation formulas among these solutions and build  transfer matrices with oscillators in the auxiliary space from the introduced degenerate Lax matrices. Finally, we provide some evidence that the constructed  transfer matrices are Baxter Q-operators for $\mathfrak{so}(2r)$ spin chains by verifying certain QQ-relations for $D_4$ at low lengths.
\end{abstract}

% \title{Oscillator construction of extremal Q-operators for orthogonal spin chains}
% 
% \title{Towards Q-operators for orthogonal spin chains}

% \title[Oscillator realisations associated to the D-type Yangian]{Oscillator realisations associated to the D-type Yangian and extremal Q-operators for orthogonal spin chains}

\title[Oscillator realisations associated to the D-type Yangian]{Oscillator realisations associated to the D-type Yangian: 
Towards the operatorial Q-system of orthogonal spin chains}

% \title[Oscillator type solutions associated to the Yangian of D-type]{Oscillator type solutions to the RTT-relation associated to the Yangian of D-type Lie algebras:\\
% Towards Q-operators for orthogonal spin chains}

\author{Rouven Frassek}

\address{
Laboratoire de Physique de l'Ecole normale supérieure, ENS - Département de Physique, 
24 rue Lhomond, 75005 Paris, France}

\maketitle
\tableofcontents

\section{Introduction}
The study of the Yangian beyond algebras of $A$-type bears many open questions. The first results on $D$-type algebras, which we will study in this note, date back to the 1970th where the fundamental R-matrix was found for the defining representation \cite{Zamolodchikov:1978xm}. Shortly after the R-matrix that intertwines the defining and the spinor representations has been obtained in \cite{Shankar:1978rb}. A few years later Reshetikhin then found the symmetric generalisations of these Lax matrices in  
 \cite{Reshetikhin:1986vd} keeping one space in the defining representation while the spinor-spinor R-matrix was presented in \cite{Reshetikhin:1986md}. Further, 
the  Bethe ansatz for $\mathfrak{so}(2r)$ has been studied in  \cite{Reshetikhin:1986vd,deVega:1986xj}. It seems that the topic of determining explicit R-matrices did not attract much attention in the following years, for which the reason may well be that closed expressions for representations corresponding to non-extremal nodes of the Dynkin diagram are difficult to obtain. Unlike in the $A$-type Yangian the irreducible representations of the Lie algebra do not lift to representations of the Yangian \cite{kirillov1990representations} and an evaluation map does not exist. See in particular \cite{MacKay:1990mp} for an explicit example of an R-matrix of this kind derived via fusion.

The Yangian arising from the R-matrix found in \cite{Zamolodchikov:1978xm} was carefully studied in the more recent work \cite{Arnaudon2006}.  In the following, we mostly follow the conventions used in that article. 
Here we also like to mention the series of papers 
\cite{Chicherin:2013rma,Chicherin:2012yn,Karakhanyan:2017wyl,Karakhanyan:2017yrg,Isaev:2017qmm,Karakhanyan:2016smw,Isaev:2015hak,karakhanyan2019spinorial} which rehabilitated the study of R-matrices and their representations  beyond $\mathfrak{sl}(r+1)$ including the ones for $\mathfrak{so}(2r)$.

In the following we will be interested in certain oscillator type realisations that were used to construct Q-operators for $A$-type algebras in \cite{Bazhanov:2010ts,Bazhanov:2010jq,Frassek:2011aa} following the ideas of \cite{Bazhanov:1998dq}. These oscillator realisations can be understood as certain degenerate limits of the Lax matrices presented in \cite{Reshetikhin:1986md}. The ordinary Yangian is obtained from the expansion of the RTT-relation in the spectral parameter  around the identity. The oscillator type solutions relevant for the construction of Q-operators do not belong to this class. The first term in the expansion of the spectral parameter can be degenerate, i.e. diagonal with vanishing matrix elements. The quantum group arising from the latter is also referred to as shifted Yangian, cf.~\cite{Braverman:2016pwk}.
We present the degenerate Lax matrices that correspond to the spinor and first fundamental representation. To obtain them we make use of the  isomorphism between $D_3$ and $A_3$ for which 
the oscillator type Lax matrices of representations apart from the first fundamental of $A$-type are known \cite{Frassek:2011aa}. The multiplicity of linear independent solutions that we find is given by the dimension of the corresponding fundamental representation.
It is also shown that the degenerate Lax matrices obey certain factorisation formulas which relate them to the Lax matrices found  in \cite{Reshetikhin:1986md}, here realised in terms of Holstein-Primakoff oscillators (free field realisation) and that they can be used to define the transfer matrices with oscillators in the auxiliary space. We expect that the transfer matrices constructed here are Q-operators and that the factorisation formulas  will allow to derive certain functional relations among them. For the case of $r=4$ we briefly address the functional relations (QQ-relations) among the Q-operators corresponding to the different nodes of the Dynkin diagram.

The paper is structured as follows. We begin in Section~\ref{sec:fundR} by discussing the relation between the R-matrix introduced in \cite{Zamolodchikov:1978xm} and the one used in \cite{Arnaudon2006} to define the Yangian. Further we introduce some symmetries of the Yang-Baxter equation. In Section~\ref{sec:so6} we discuss the isomorphism between $A_3$ and $D_3$ and use the degenerate Lax matrices obtained in \cite{Frassek:2011aa} to derive the ones for $D_3$. In Section~\ref{sec:minimal} we show that the Lax matrices corresponding to the spinor nodes of the Dynkin diagram can be lifted to $D_r$ and show that they satisfy the Yang-Baxter equation. We also give the Lax matrices at the first fundamental node for $D_r$. In Section~\ref{sec:factorisation} we discuss several factorisation formulas and Section~\ref{sec:Qops} is devoted to the construction of transfer matrices/Q-operators using some of the previously derived Lax matrices and also contains some QQ-relations for $r=4$.  Finally in Section~\ref{sec:conc} we end with a conclusion. 

\section{Fundamental R-matrix}\label{sec:fundR}

The fundamental R-matrix for $\mathfrak{so}(2r)$ corresponding to the first fundamental representation was obtained in  \cite{Zamolodchikov:1978xm}. It is a $4r^2\times 4r^2$ matrix acting on the tensor product of two spaces $\mathbb{C}^{2r}\otimes \mathbb{C}^{2r}$ and is  commonly written as
\begin{equation}\label{eq:normalR}
 R(z)=z(z+\ka)\ID +(z+\ka)\Pop -z\Kop \,.
\end{equation}  
Here $z$ is the spectral parameter and $\ka$ is related to the rank of the Lie algebra via $\ka=r-1$. Furthermore we  introduced  the identity $\id$,  the permutation $\Pop $ and the  $\Kop $ matrix. The latter ones are defined as
\begin{equation}
%  \ID =\sum_{A=1}^{n}e_{AA}\otimes e_{AA}\,,\qquad  
 \Pop =\sum_{A,B=1}^{2r}e_{AB}\otimes e_{BA}\,,\qquad  \Kop =\sum_{A,B=1}^{2r}e_{AB}\otimes e_{AB}\,,
\end{equation} 
respectively. Here $e_{AB}$ denotes the  $2r\times 2r$  unit matrices with $\left(e_{AB}\right)_{CD}=\delta_{AC}\delta_{BD}$.
Throughout this article  we  work in the basis used in \cite{Arnaudon2006}.  In this basis the R-matrix is written as
\begin{equation}\label{eq:molevR}
 \RR(z)= z(z+\ka)\ID+(z+\ka)\Pop -z\,\Qop \,,
\end{equation} 
where   $\Pop $ again denotes the permutation operator but  $\Qop $ differs from $ \Kop $. We have
\begin{equation}
 \Pop =\sum_{\substack{a,b=-r\\a,b\neq0}}^{r}E_{ab}\otimes E_{ba}\,,\qquad  \Qop =\sum_{\substack{a,b=-r\\a,b\neq0}}^{r}E_{a,b}\otimes E_{-a,-b}\,.
\end{equation} 
The unit matrices   $E_{ab}$ denote $2r\times 2r$ matrices with the indices $a,b\in\{-r,\ldots,-1,1,\ldots,r\}$. They can be  defined in terms of the unit matrices $e_{AB}$, with $A,B\in\{1,\ldots,2r\}$,  via
\begin{align}\label{eq:defE1}
    E_{-i,-j}&= e_{r-i+1,r-j+1} \,,         &  E_{-i,j} &= e_{r-i+1,r+j} \,, \\ 
 E_{i,-j}&=  e_{r+i,r-j+1}\,,  &  E_{i,j}  &= e_{r+i,r+j}\,.\label{eq:defE2}
\end{align}
The small Latin indices take the values  $i,j\in\{1,\ldots,r\}$. The explicit relation between the R-matrices  \eqref{eq:normalR} and \eqref{eq:molevR} reads
% is  are related by a similarity transformation
\begin{equation}\label{eq:transr}
 (\Sop\otimes\Sop)\R(z)  (\Sop^{-1}\otimes\Sop^{-1})=\RR(z)\,,
\end{equation} 
where the similarity transformation is given by the $2r\times 2r$ block matrices 
\begin{equation}\label{eq:simtrans}
 \Sop=\frac{1}{\sqrt{2}}\left(\begin{BMAT}[5pt]{c:c}{c:c}
      -i\, \idb &\idb  \\
      i \, \ID &\ID 
      \end{BMAT}
\right)\,,\qquad 
 \Sop^{-1}=\frac{1}{\sqrt{2}}\left(\begin{BMAT}[5pt]{c:c}{c:c}
      i \,\idb &-i\,\id\\
       \idb &\id
      \end{BMAT}
\right)\,,
\end{equation} 
with $i^2=-1$.
Each block in \eqref{eq:simtrans} is of the size $r\times r$ and $\id$ denotes the identity matrix of appropriate size while $\idb$ denotes the exchange matrix or  reversed identity matrix defined as
\begin{equation}\label{eq:invar}
\idb =\left(\begin{array}{ccc}
                         0&0&1\\
                         0&\iddots&0\\
                         1&0&0\\
                        \end{array}
\right)\,.
\end{equation} 
The relation above in \eqref{eq:transr} can be verified noting that the identity and the permutation matrix are unaffected by the similarity transformations while $
 (\Sop\otimes\Sop)\Kop  (\Sop^{-1}\otimes\Sop^{-1})=\Qop $ as shown in Appendix~\ref{app:gens}.
 
 In the following  we study solutions to the RTT-relation 
\begin{equation}\label{eq:rtt}
 \RR(x-y)\left(\L(x)\otimes\ID \right)\left(\ID \otimes \L(y))\right)=\left(\ID \otimes \L(y)\right)\left(\L(x)\otimes\ID \right)\RR(x-y)\,,
\end{equation} 
where $\L$ denotes a $2r\times 2r$ matrix with non-commuting entries. Any solution $\L$ of the RTT-relation \eqref{eq:rtt} can be transformed via 
$\Sop^{-1}\L(z)\Sop$ to yield the corresponding solution to the RTT-relation with $\RR(x-y)$ exchanged by $\R(x-y)$.
As discussed in \cite{Arnaudon2006} the R-matrix \eqref{eq:molevR} is invariant under the transformation
\begin{equation}\label{eq:inv}
 [\RR(z),\Bop\otimes \Bop]=0\,,
\end{equation} 
if $\Bop\Bop'=\id$. Here the prime denotes the transposition $E_{ab}'=E_{-b,-a}$. This symmetry stems from the $\mathfrak{so}(2r)$ invariance of the R-matrix $R$ in \eqref{eq:normalR}. It follows that $\Bop\,\L(z)\, \Bop'$ is a solution to the RTT-relation \eqref{eq:rtt} if $\L(z)$ is.

Finally, we introduce two sets of permutation matrices that satisfy the invariance condition \eqref{eq:inv}. The first one is labelled by a vector $\vec\alpha=(\alpha_1,\alpha_2,\ldots,\alpha_r)$ with elements $\alpha_i=\pm 1$. When applied to $\L$ it   permutes the $i$th and $-i$th row and $i$th and $-i$th column for each $i$ with $\alpha_i=-1$. It can explicitly be realised as
\begin{equation}\label{eq:invtrans}
 B(\vec\alpha)=\frac{1}{2}\sum_{i=1}^r\left((1+\alpha_i)(E_{-i,-i}+E_{i,i})+(1-\alpha_i)(E_{-i,i}+E_{i,-i})\right)\,.
\end{equation} 
The second permutation matrix is labelled by two indices $i,j=1,\ldots r$. When applied to $\L$ it   permutes the $i$th and $j$th rows and columns and  the $-i$th and $-j$th rows and columns. It can be written as
\begin{equation}\label{eq:invtrans2}
\tilde B_{ij}=\sum_{\substack{k=1\\k\neq i,j}}^r (E_{-k,-k}+E_{k,k})+E_{-i,-j}+E_{-j,-i}+E_{i,j}+E_{j,i}\,.
\end{equation} 
Here the matrix $\tilde B_{ij}$ is symmetric in the exchange of $i$ and $j$ and we assumed that $1\leq i\neq j\leq r$.
It is straightforward to show that the matrices above satisfy the conditions $B(\vec\alpha)=B'(\vec\alpha)$ and $\tilde B_{ij}=\tilde B'_{ij}$ as well as the conditions $\Bop\Bop'=\ID$ . For the matrices $B(\vec \alpha)$ we find that
\begin{equation}
 B(\vec\alpha) B'(\vec\alpha)=\frac{1}{2}\sum_{i=1}^r\left((1+\alpha_i^2)(E_{i,i}+E_{-i,-i})+(1+\alpha_i)(1-\alpha_i)(E_{-i,i}+E_{i,-i})\right)=\ID\,,
\end{equation} 
where we used that $\alpha_i=\pm1$ in the last step and for the matrices $\tilde B_{ij}$ we get
\begin{equation}
 \tilde B_{ij}\tilde B'_{ij}=\sum_{\substack{k=1\\k\neq i,j}}^r (E_{-k,-k}+E_{k,k})+E_{-i,-i}+E_{-j,-j}+E_{i,i}+E_{j,j}=\id\,.
\end{equation}

\section{Lax matrices for $D_3$}\label{sec:so6}
\newcommand{\quuz}{
\begin{tikzpicture}
\foreach \a in {1,2} {
    \begin{scope}[shift={(0.7*\a,0)}]
      \draw[fill=lightgray!30] (0.3*\a,0) circle (0.3cm);
      \draw[black,thick] (0.3*\a+0.3,0)--(0.3*\a+0.7,0);
    \end{scope}
  }
      \draw[fill=lightgray!30] (3,0) circle (0.3cm);
  \node at (1,0) {$4$};
  \node at (2,0) {$6$};
  \node at (3,0) {$4$};
       \node [left] at (0,0) {$A_3:$};
\end{tikzpicture}
}
\begin{figure}
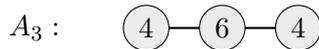

\begin{center}
 \quuz
\end{center}
 \caption{Dynkin diagram for $A_3$ with the number of degenerate Lax-operators corresponding to Q-operators indicated for each node.}
 \label{fig:su3}
\end{figure}

Before studying solutions to the RTT-relation \eqref{eq:rtt} for general rank we restrict ourselves to $r=3$.  In this case we can obtain the Lax matrices for Q-operators  by making use of the isomorphism between the algebras $A_3$ and $D_3$. We keep the presentation short as this section only serves to motivate the Lax matrices introduced for $\mathfrak{so}(2r)$ in Section~\ref{sec:minimal}.

The relevant degenerate Lax matrices for $A_{r}$ of oscillator type were  derived in \cite{Frassek:2011aa}. In the case of $A_3$ there are in total $16$  such solutions $L^{rep}_I(z)$  labelled by the set $I$ which takes values $I\subseteq \{1,2,3,4\}$ and the representation (``$rep$'') in the matrix space. Here the full and the empty set represent trivial solutions which are proportional to the identity. The solutions can be distinguished by the level $|I|=k$ with $k=0,\ldots,4$ which corresponds up to the trivial solutions to the $k$th node on the Dynkin diagram of $A_3$.  At each level there are $\binom{4}{k}$ solutions as indicated in Figure~\ref{fig:su3}. They can be evaluated straightforwardly once the representation of the algebra is fixed.
Here we evaluate the degenerate Lax matrices for the antisymmetric six-dimensional representation '$\mathbf{6}$' of $A_3$.
Following the notations used in \cite{Frassek:2011aa}  we find that the level $1$ and level $3$ Lax matrices are linear in the spectral parameter. In particular for the set $I=\{1,2,3\}$ the Lax matrix can be brought to the form
\begin{equation}\label{eq:linlaxso6}
 {\eL}_{\{1,2,3\}}(z)=\left(\begin{array}{cccccc}
z-\NN_1-\NN_2&-\oad_1\oa_3&\oad_2\oa_3&\oad_1&\oad_2&0\\
-\oad_3\oa_1&z-\NN_2-\NN_3&-\oad_2\oa_1&\oad_3&0&-\oad_2\\
\oad_3\oa_2&-\oad_1\oa_2&z-\NN_1-\NN_3&0&-\oad_3&-\oad_1\\
-\oa_1&-\oa_3&0&1&0&0\\
-\oa_2&0&\oa_3&0&1&0\\
0&\oa_2&\oa_1&0&0&1\\
              \end{array}\right)\,,
\end{equation} 
where we introduced the oscillators $[\oa_i,\oad_j]=\delta_{ij}$.
One can check that this Lax matrix is a solution to the RTT-relation \eqref{eq:rtt}. Further, by evaluating all other remaining Lax matrices of the $1$st and $3$rd level one finds that, up to particle hole transformations and renaming of the oscillators, all these solutions can be brought to the form $B(\vec\alpha) {\eL}_{\{1,2,3\}}(z)B'(\vec\alpha)$  with the similarity transformation given in  \eqref{eq:invtrans} for some choice of $\vec\alpha$. Here we observe that the similarity transformations with $\prod_{i=1}^3\alpha_i=1$ map to the degenerate Lax operators at the same level $3$, i.e. the same note on the Dynkin diagram, while the ones with $\prod_{i=1}^3\alpha_i=-1$ map to the opposite extremal node corresponding to level $1$. 

Similarly we can evaluate the $2$nd level degenerate Lax matrices and find that the solutions are quadratic in the spectral parameter. In particular we find that for the set $I=\{1,2\}$ they can be brought to the form
\begin{equation}\label{eq:r12}
{\eL}_{\{1,2\}}(z)=e^{M_+}M_0(z)e^{M_-}\,,
\end{equation} 
with
\begin{equation}
 M_+=\left(\begin{array}{cccccc}
0&\oad_1&\oad_2&\oad_3&\oad_4&0\\
0&0&0&0&0&-\oad_4\\
0&0&0&0&0&-\oad_3\\
0&0&0&0&0&-\oad_2\\
0&0&0&0&0&-\oad_1\\
0&0&0&0&0&0\\
              \end{array}\right)\,,\quad 
 M_-=\left(\begin{array}{cccccc}
0&0&0&0&0&0\\
-\oa_1&0&0&0&0&0\\
-\oa_2&0&0&0&0&0\\
-\oa_3&0&0&0&0&0\\
-\oa_4&0&0&0&0&0\\
0&\oa_4&\oa_3&\oa_2&\oa_1&0\\
              \end{array}\right)\,,
\end{equation} 
and the middle part
\begin{equation}
M_0(z)=\diag(z^2-z,z,z,z,z,1)\,.
\end{equation}
Again one can check that the Lax matrix $ {\eL}_{\{1,2\}}(z)$ in \eqref{eq:r12} is a solution to the RTT-relation \eqref{eq:rtt}. The remaining degenerate LAx matrices of the $2$nd level can be related  to the R-matrix  ${\eL}_{\{1,2\}}(z)$, again up to particle-hole transformations and renaming of the oscillators, by applying the the similarity transformations as defined in \eqref{eq:invtrans2} and the similarity transformation  with $B(-1,-1,-1)$. 
 
The solutions of the corresponding YBE with $R(z)$ in \eqref{eq:normalR} can be obtained through the similarity transform \eqref{eq:simtrans}. However, we note that in this basis the spectral parameter does not only appear on the diagonal in the Lax matrix \eqref{eq:linlaxso6} and not only on the diagonal of the middle part ${\eL}_0(z)$ of the Lax matrix \eqref{eq:r12}.

\section{Minimal solutions of oscillator type for $D_r$}\label{sec:minimal}
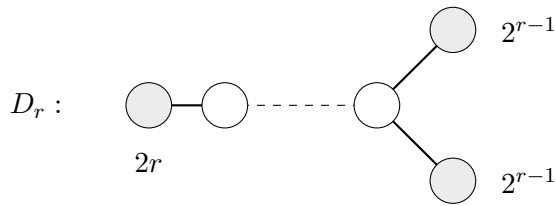
\begin{figure}
\begin{center}
 \begin{tikzpicture}
\foreach \a in {1} {
    \begin{scope}[shift={(0.7*\a,0)}]
      \draw[fill=lightgray!30] (0.3*\a,0) circle (0.3cm);
      \draw[black,thick] (0.3*\a+0.3,0)--(0.3*\a+0.7,0);
    \end{scope}
  }
      \draw[black,dashed] (2,0)--(4,0);
      \draw[black,thick] (4,0)--(5,1);
      \draw[black,thick] (4,0)--(5,-1);
      \draw[fill=white] (2,0) circle (0.3cm);
      \draw[fill=white] (4,0) circle (0.3cm);
      \draw[fill=lightgray!30] (5,1) circle (0.3cm);
      \draw[fill=lightgray!30] (5,-1) circle (0.3cm);
       \node [left] at (0,0) {$D_r:$};
       \node [below] at (1,-0.5) {$2r$};
       \node [right] at (5.5,-1) {$2^{r-1}$};
       \node [right] at (5.5,1) {$2^{r-1}$};
\end{tikzpicture}
\end{center}
\caption{Dynkin diagram for $D_r$ with the expected number of R-operators corresponding to the dimension of the fundamental representations on the extremal nodes.}
\label{fig:dr}
\end{figure}
In this section we introduce a set of new solutions of oscillator type and
% As it is rather cumbersome to actually derive the solutions w
show that they satisfy the RTT-relation \eqref{eq:rtt}.\footnote{Parts of the discussion are postponed to Section~\ref{sec:spinf}.} To do so it is convenient to introduce the $2r\times 2r$  matrix
\begin{equation}
 \L(x)=\sum_{a,b=-r}^{r}\L_{a,b}(x)E_{ab}\,,
\end{equation} 
with non-commuting entries $\L_{a,b}(x)$
and write the RTT-relation as the commutation relations
\begin{equation}\label{eq:yangian}
\begin{split}
 [\L_{a,b}(x),\L_{c,d}(y)]&=\frac{1}{x-y}\left(\L_{c,b}(y)\L_{a,d}(x)-\L_{c,b}(x)\L_{a,d}(y)\right)\\
 &\quad+\frac{1}{x-y+\kappa}\left(\delta_{a,-c}\left(\L^t(x)\,\idb \,\L(y)\right)_{bd}-\delta_{b,-d}\left(\L(y)\,\idb \,\L^t(x)\right)_{ca}\,\right)\,,
 \end{split}
\end{equation} 
cf.~\cite{Arnaudon2006}. Here we defined the transposition as $E_{ab}^t=E_{ba}$ and $\idb$ was defined in \eqref{eq:invar}.

In the following we present one family of solutions that is linear in the spectral parameter and one that is quadratic. They correspond to the spinor nodes and the first fundamental node respectively, cf.~Figure~\ref{fig:dr}. 
Their form in the basis of $R$ in \eqref{eq:normalR} can be obtained through the similarity transform \eqref{eq:simtrans}.

\subsection{Linear solutions}\label{sec:linsol}
The first Lax matrix we introduce is linear in the spectral parameter and contains $\frac{r(r-1)}{2}$ pairs oscillators. It is the generalisation of \eqref{eq:linlaxso6} and can conveniently be written as a $2\times 2$ block matrix
\begin{equation}\label{eq:laxlin}
 \Lspi(z)=
\left(\begin{BMAT}[5pt]{c:c}{c:c}
      z\,\id+\ap\am&\ap\\
    \am&\id
      \end{BMAT}
\right)\,,
\end{equation} 
where each block is of the size $r\times r$ and $\id$ denotes the identity matrix. Explicitly the matrices $\ap$ and $\am$ which contain the oscillators $[\oa_{i,-j},\oad_{-k,l}]=\delta_{il}\delta_{jk}$ are of the form
\begin{equation}\label{eq:ApAm}
\ap=\left(\begin{array}{cccc}
               \oad_{-r,1}&\cdots&\oad_{-r,r-1}&0\\
               \vdots  &\iddots &0&-\oad_{-r,r-1}\\
             \oad_{-2,1}&0&\iddots&\vdots\\
               0&-\oad_{-2,1}&\cdots&-\oad_{-r,1}
              \end{array}\right),\;
              \am=\left(\begin{array}{cccc}
              - \oa_{1,-r}&\cdots&-\oa_{1,-2}&0\\
              \vdots&\iddots &0&\oa_{1,-2}\\
               -\oa_{r-1,-r}&0&\iddots&\vdots\\
               0&\oa_{r-1,-r}&\cdots&\oa_{1,-r}
              \end{array}\right).
\end{equation} 
Alternatively, the Lax matrix above can naturally be written in the factorised form 
\begin{equation}\label{eq:spinfac}
 \Lspi(z)=
\left(\begin{BMAT}[5pt]{c:c}{c:c}
      \id&\ap\\
    0&\id
      \end{BMAT}
\right)
\left(\begin{BMAT}[5pt]{c:c}{c:c}
      z\,\id&0\\
    0&\id
      \end{BMAT}
\right)
\left(\begin{BMAT}[5pt]{c:c}{c:c}
      \id&0\\
    \am&\id
      \end{BMAT}
\right)\,.
\end{equation} 
In the following we show that the Lax matrix $\Lspi(z)$ in \eqref{eq:laxlin} satisfies the commutation relations of the Yangian in \eqref{eq:yangian} and therefore is a solution to the RTT-relation \eqref{eq:rtt}.

We begin with computing the explicit form of the products in the second line of \eqref{eq:yangian}. One finds
\begin{equation}\label{eq:LtL}
 \Lspi^t(x)\,\idb \,\Lspi(y)=
\left(\begin{BMAT}[5pt]{c:c}{c:c}
      (x-y+\ka)\idb \am&(x+\ka)\idb \\
     y\idb &0
      \end{BMAT}
\right)\,,\quad  \Lspi(y)\,\idb \,\Lspi^t(x)=
\left(\begin{BMAT}[5pt]{c:c}{c:c}
      (x-y+\ka)\ap\idb &y\idb \\
     (x+\ka)\idb &0
      \end{BMAT}
\right)\,.
\end{equation} 
Here we used the relations $\idb\,\ap\,\idb=-\ap^T$ and $\idb\,\am\,\idb=-\am^T$ where $\ap^T$ and $\am^T$ denote the transpose of the matrices $\ap$ and $\am$ in \eqref{eq:ApAm} along the main diagonal.

Further it  is convenient to split the commutation relations \eqref{eq:yangian} that arise from the RTT-equation  into block structure and write the resulting equations as a $4\times 4$ block matrix, cf.~\eqref{eq:laxlin}. 
We note that for the choice of the Lax operator \eqref{eq:laxlin}  the first and second term on the the right-hand-side of \eqref{eq:yangian} are antisymmetric 
under the consecutive exchange:  $a\leftrightarrow c$,  $b\leftrightarrow d$ and $x\leftrightarrow y$.  This is consistent with the commutator on the left-hand-side. Thus the $4 \times 4$ matrix is antisymmetric and the number of constraining equations reduces  to $10$.
In the following we  verify these  commutation relations. 

We begin with the diagonal terms. Obviously the term in the second line of \eqref{eq:yangian} does not contribute  and the only non-vanishing commutator arises for   $a=-i$, $b=-j$, $c=-k$ and $d=-l$. It reads
\begin{equation}
\begin{split}
 [(\ap\am)_{-i,-j},(\ap\am)_{-k,-l}]&=(\ap\am)_{-k,-j}\delta_{i,l}-\delta_{k,j}(\ap\am)_{-i,-l}\,,
 \end{split}
\end{equation} 
and can be verified using the explicit form of $\ap$ and $\am$ in \eqref{eq:ApAm}.
Here and in the following we use the notation $\ap_{-i,j}$  to denote the element in the $(r-i+1)$th row and $j$th column of the matrices in $\ap$ and  $\am_{i,-j}$  to denote the element in the $i$th row and $(r-j+1)$th column of the matrices in $\am$ in \eqref{eq:ApAm}, cf.~\eqref{eq:defE1} and \eqref{eq:defE2}. Such that $\ap_{-i,j}=-\ap_{-j,i}$ and $\am_{i,-j}=\am_{j,-i}$ and 
\begin{equation}\label{eq:osccom}
[\am_{i,-j},\ap_{-k,l}]= \delta_{ik}\delta_{jl}- \delta_{il}\delta_{kj}\,,
\end{equation} 
where $i,j,k,l\in\{1,\ldots,r\}$.
We verify the remaining $6$ conditions case by case. For $a=-i$, $b=-j$, $c=-k$, $d=l$ we get 
\begin{equation}
[(\ap\am)_{-i,-j},\ap_{-k,l}]=\ap_{-i,k} \delta_{jl}-\ap_{-i,l}\delta_{jk}\,,
\end{equation} 
which can be shown straightforwardly employing \eqref{eq:osccom}. Here we used that   $(\ap\idb)_{-k,-i}=-\ap_{-i,k}$. Similarly, using $(\idb\am)_{-j,-l}=-\am_{l,-j}$ we find 
\begin{equation}
% [(\ap\am)_{-i,-j},\ap_{-k,l}]=\ap_{-i,k} \delta_{jl}-\ap_{-i,l}\delta_{jk}\,,\qquad 
[(\ap\am)_{-i,-j},\am_{k,-l}]=\am_{k,-j} \delta_{il}-\am_{l,-j}\delta_{ik}\,,
\end{equation} 
  for $a=-i$, $b=-j$, $c=k$, $d=-l$  which follows from \eqref{eq:osccom}. Further the case $a=-i$, $b=j$, $c=k$, $d=-l$ reduces to \eqref{eq:osccom}. Here the first and second term on the right-hand-side stem from the  first and second term in \eqref{eq:yangian} respectively.  For the remaining cases one has to verify that the second term on the right-hand-side of \eqref{eq:yangian} vanishes. This can be done using the explicit expression given in \eqref{eq:LtL}.
   Finally we conclude that the Lax matrix \eqref{eq:laxlin} is a solution to the RTT-relation \eqref{eq:rtt}.

We note that one can  obtain further solutions by permuting the $-i$th and $i$th row and column using the transformation in \eqref{eq:invtrans}. In this way we obtain solutions with a different distribution of the  spectral parameter on the diagonal.  In total we find $2^r$  solution
\begin{equation}
\Lspi_{\vec\alpha}(z)=B(\vec\alpha)\Lspi(z)B(\vec\alpha)\,,
\end{equation} 
 which are  labelled by the vector $\vec\alpha=(\alpha_1,\ldots,\alpha_r)$ with $\alpha_i=\pm1$.

\subsection{Quadratic  solutions}
The second Lax matrix we introduce is quadratic in the spectral parameter and contains $2(r-1)$ pairs of oscillators. It generalises the Lax matrix for $D_3$ in \eqref{eq:r12} and can be written as the block matrix
\begin{equation}\label{eq:quadlax}
L(z)=
 \left(\begin{BMAT}[5pt]{c|c|c}{c|c|c}
z^2+z(2-r-\wp\wm)+\frac{1}{4}\wp\idb\wp^t\wm^t\idb\wm&z\wp-\frac{1}{2}\wp\idb\wp^t\wm^t\idb&-\frac{1}{2}\wp\idb\wp^t\\\
-z\wm+\frac{1}{2}\idb\wp^t\wm^t\idb\wm&z\id-\idb\wp^t\wm^t\idb&-\idb\wp^t\\
-\frac{1}{2}\wm^t\idb\wm&\wm^t\idb&1\\
      \end{BMAT}
\right)\,.
\end{equation}
Here the size of the blocks on the diagonal is $1\times 1$, $(2r-2)\times(2r-2)$ and $1\times1$, respectively. Further we defined the vectors $\wp$ and $\wm$ containing the oscillators $[\oa_a,\oad_b]=\delta_{ab}$ with the indices taking the values $a,b=-r+1,\ldots,-1,1,\dots,r-1$ as
\begin{equation}
 \wp=(\oad_{-r+1},\ldots,\oad_{-1},\oad_{1},\ldots,\oad_{r-1})\,,\qquad \wm=(\oa_{-r+1},\ldots,\oa_{-1},\oa_1,\ldots,\oa_{r-1})^t\,.
\end{equation}
It follows that 
\begin{equation}\label{eq:vecs}
 \frac{1}{2}\wp^t\idb\wp=\sum_{k=1}^{r-1}\oad_{k}\oad_{-k}\,,\qquad 
 \frac{1}{2}\wm^t\idb\wm=\sum_{k=1}^{r-1}\oa_{k}\oa_{-k}\,.
\end{equation} 
We note that it  is rather involved to explicitly check that the Lax matrix in \eqref{eq:quadlax} satisfies the RTT-relation \eqref{eq:rtt}. However we were able to verify it for $r= 4$. An argument for arbitrary $r$ based on the factorisation of two spinor type Lax matrices \eqref{eq:laxlin}  is provided  in Section~\ref{sec:spinf}. 

By acting with the similarity transformation \eqref{eq:invtrans2} we obtain $r$ solutions with different distributions of the spectral parameter on the diagonal. All these solutions have the term quadratic in the spectral parameter on a different position on the diagonal of the upper left $r\times r $ block matrix. Using the similarity transformation \eqref{eq:invtrans} with $\alpha_i=-1$ we obtain the corresponding solutions with the quadratic term  on the diagonal of the lower right $r\times r $ block matrix. In total we find $2r$ solutions.

Finally we note that the Lax matrix in \eqref{eq:quadlax}  can be written in the factorised form

\begin{equation}\label{eq:fundfac}
L(z)=
 \left(\begin{BMAT}[5pt]{c|c|c}{c|c|c}
1&\wp&-\frac{1}{2}\wp\idb\wp^t\\\
0&\id&-\idb \wp^t\\
0&0&1\\
      \end{BMAT}
\right)
 \left(\begin{BMAT}[5pt]{c|c|c}{c|c|c}
z(z-r+2)&0&0\\\
0&z\id&0\\
0&0&1\\
      \end{BMAT}
\right)
 \left(\begin{BMAT}[5pt]{c|c|c}{c|c|c}
1&0&0\\
-\wm&\id&0\\
-\frac{1}{2}\wm^t\idb\wm&\wm^t\idb&1\\
      \end{BMAT}
\right)\,.
\end{equation}
A similar factorisation as for the Lax matrices studied here for $D_r$ in \eqref{eq:spinfac} and \eqref{eq:fundfac} was observed in \cite{Frassek:2011aa} for the case of $A_r$.

\section{Factorisation formulas}\label{sec:factorisation}
In the following we study certain factorisation formulas among the degenerate Lax matrices that we have introduced in the previous section. More precisely we derive the oscillator realisation of the Lax matrices for the spinor and first fundamental representation in Section~\ref{sec:spinlax} and \ref{sec:fundlax} respectively. In other words we recover the Holstein-Primakoff realisation (free field realisation) of the $\mathfrak{so}(2r)$ generators $F_{a,b}$ with $F_{ab}=-F_{-b,-a}$ satisfying the commutation relations
\begin{equation}\label{eq:comrel}
 [F_{ab},F_{cd}]=\delta_{c,b}F_{a,d}-\delta_{a,d}F_{c,b}-\delta_{c,-a}F_{-b,d}+\delta_{d,-b}F_{c,-a}\,.
\end{equation}

Additionally, in Section~\ref{sec:spinf} we present a relation between the linear and quadratic solutions in Section~\ref{sec:minimal} which may serve as a proof, which we had skipped earlier, that the quadratic Lax matrix in  \eqref{eq:quadlax} indeed satisfies the RTT-relation.

\subsection{Spinor representation}\label{sec:spinlax}
In this subsection we derive the full Lax matrix in oscillator form corresponding to a spinor representation. We introduce the Lax matrix
\begin{equation}
 \Lspi_{(-1,\ldots,-1)}(z)=
\left(\begin{BMAT}[5pt]{c:c}{c:c}
 \id&-\am^T\\
    -\ap^T&     z\,\id+\ap^T\am^T
      \end{BMAT}
\right)\,,
\end{equation} 
which can be obtained by applying the similarity transformation \eqref{eq:invtrans} to \eqref{eq:laxlin} and thus satisfies the RTT-relation.
Multiplying this solution with $\Lspi(z)$ defined in \eqref{eq:laxlin} in the matrix space and taking the tensor product in the oscillator space we find that the result can be decomposed as
\begin{equation}\label{eq:facform1}
  \Lspi^{[1]}(z+s) \Lspi^{[2]}_{(-1,\ldots,-1)}(z-s-\kappa)=\mathcal{S}\mathcal{L}^{[1]}_s(z)\Gop ^{[2]}\mathcal{S}^{-1}\,.
\end{equation} 
Here $\mathcal{L}_s$ denotes the Lax matrix
\begin{equation}\label{eq:laxspin}
 \mathcal{L}_s^{[1]}(z)=\left(\begin{BMAT}[5pt]{c:c}{c:c}
(z+s) \id +\ap_{[1]}\am_{[1]}&-\ap_{[1]}(2s+\kappa+\am_{[1]}\ap_{[1]})\\
    \am_{[1]}&   (z-s-\kappa) \id-\am_{[1]}\ap_{[1]}
      \end{BMAT}
\right)\,,
\end{equation} 
while 
\begin{equation}
 \Gop ^{[2]}=\left(\begin{BMAT}[5pt]{c:c}{c:c}
 \id&-\am^T_{[2]}\\
    0&     \id
      \end{BMAT}
\right)\,,
\end{equation} 
is independent of the spectral parameter. It is straightforward to verify that the matrix $\Gop$ satisfies $\Gop\Gop'=\ID$ and thus solves the RTT-relation. 
The similarity transform $\mathcal{S}$ is defined via
\begin{equation}
 \mathcal{S}=\exp [-\tr(\ap_{[1]}\ap^T_{[2]})]\,,
\end{equation} 
induces the shifts $ \mathcal{S}\am_{[1]}\mathcal{S}^{-1}=\am_{[1]}-\ap^T_{[2]}$ and $\mathcal{S}\am^T_{[2]}\mathcal{S}^{-1}=\am^T_{[2]}-\ap_{[1]}$. It is introduced to separates the dependence in the oscillators. We remark that this factorisation formula is similar to the one in the $\mathfrak{su}(2)$ case studied in \cite{Bazhanov:2010ts}. It is clear that the product on the left-hand-side of  \eqref{eq:facform1} satisfies the RTT-relation. This is a consequence of the computation done in Section~\ref{sec:linsol} and the symmetries of the R-matrix as discussed in Section~\ref{sec:fundR}. Therefore also the expression on the right-hand-side satisfies the RTT-relation. Further, as $\mathcal{S}$ only depends on the oscillators and $\Gop$, whose entries commute with the ones of $\mathcal{L}_s$, satisfies the RTT-relation we can conclude that also $\mathcal{L}_s$ satisfies the RTT-relation. A similar argument has been given in  \cite{Bazhanov:2010ts}.

The Lax matrix in \eqref{eq:laxspin} is linear in the spectral parameter and thus we expect that the oscillators realise a spinor representation of $D_r$, cf.~\cite{Shankar:1978rb,Reshetikhin:1986md}. This can be seen when identifying the generators via
\begin{equation}\label{eq:slax}
 \mathcal{L}_s(z)=z+\sum_{a,b}E_{ab}F^s_{ba}\,.
\end{equation} 
The generators  $F_{ab}^s$ defined by equating \eqref{eq:laxspin}   and  \eqref{eq:slax} satisfy the commutation relations in \eqref{eq:comrel}. Further, when acting on the Fock vacuum which is annihilated by $\oa_{ab}$ such that $\oa_{ab}|0\rangle=0$ for arbitrary $a$ and $b$ we find
\begin{equation}
 F_{ab}|0\rangle =0 \quad \text{for} \quad a<b\,,\qquad 
 F_{ii}|0\rangle =-f_i|0\rangle \,.
\end{equation} 
The weight vector $f=(f_1,\ldots,f_r)$ is given by $f=(s,\ldots,s)$.
Thus the generators defined in this way indeed realise a spinor representation with the Fock vacuum $|0\rangle$ as the highest weight state, cf.~e.g. \cite{Das2014}. This representation is finite-dimensional for $2s\in\mathbb{N}_0$. 
The generators obey the characteristic identity
\begin{equation}
\left(F_{ab}^s+s\delta_{ab}\right)\left(F_{bc}^s-(s+\kappa)\delta_{bc}\right)=0\,.
%  \sum_c F_{ac}^sF_{cb}^s=\kappa F_{ab}^s+s(s+\kappa)
\end{equation} 
This can be verified using the factorised form of the Lax matrix\,\footnote{
A similar Lax matrix in factorised form was obtained in \cite{Chicherin:2012yn} where the matrix space is given by a  spinor representation. }
\begin{equation}
 \mathcal{L}_s^{[1]}(z)=\left(\begin{BMAT}[5pt]{c:c}{c:c}
 \id&\ap_{[1]}\\
    0&     \id
      \end{BMAT}
\right)\left(\begin{BMAT}[5pt]{c:c}{c:c}
(z+s) \id&0\\
    \am_{[1]}&   (z-s-\kappa) \id
      \end{BMAT}
\right)\left(\begin{BMAT}[5pt]{c:c}{c:c}
 \id&-\ap_{[1]}\\
    0&    \id      \end{BMAT}
\right)\,.
\end{equation} 
Here the difference in sign in the characteristic identity compared to \cite{Reshetikhin:1986vd} arises from the definition of the generators. 

The Lax matrix \eqref{eq:laxspin} corresponding to the  spinor representation $f=(-s,s,\ldots,s)$ can be obtained by applying the similarity transformation \eqref{eq:invtrans} with $\alpha_1=-1$ and $\alpha_i=1$ for $i=2,3,\ldots,r$ to the Lax matrix in \eqref{eq:laxspin}  in order to exchange the rows and columns corresponding to the indices $1$ and $-1$.
Further we note that the Lax matrix \eqref{eq:laxlin} can be deduced from \eqref{eq:laxspin} by taking a limit
\begin{equation}
\Lspi(z)=\lim_{s \to \infty } \frac{\mathcal{L}_s(z-s)}{i\sqrt{2s}} \left(\begin{BMAT}[5pt]{c:c}{c:c}
i\sqrt{2s}\,\ID&0\\
   0& -\frac{i}{\sqrt{2s}}\,\id
      \end{BMAT}
\right)\,,
\end{equation} 
in analogy to the $\mathfrak{sl}(2)$ case discussed in e.g. \cite{sklyanin00,Bazhanov:2010ts}. Here the matrix on the right satisfies the condition in \eqref{eq:inv}.
% 
% 
% Finally we present the Lax matrix \eqref{eq:laxspin} in  factorised form. It reads
% \begin{equation}
%  \mathcal{L}_s^{[1]}(z)=\left(\begin{BMAT}[5pt]{c:c}{c:c}
%  \id&\ap_{[1]}\\
%     0&     \id
%       \end{BMAT}
% \right)\left(\begin{BMAT}[5pt]{c:c}{c:c}
% (z+s) \id&0\\
%     \am_{[1]}&   (z-s-\kappa) \id
%       \end{BMAT}
% \right)\left(\begin{BMAT}[5pt]{c:c}{c:c}
%  \id&-\ap_{[1]}\\
%     0&    \id      \end{BMAT}
% \right)\,.
% \end{equation} 

\subsection{From spinor to first fundamental}\label{sec:spinf}
We will now derive the quadratic Lax matrices \eqref{eq:quadlax} from the linear Lax matrices \eqref{eq:laxlin}.
For this purpose it is convenient to write the linear Lax matrix in \eqref{eq:laxlin} as a $4\times 4$ block matrix
\begin{equation}\label{eq:lindec}
\Lspi(z)=
 \left(\begin{BMAT}[5pt]{c|c:c|c}{c|c:c|c}
z-\vp\vm&\vp\am&\vp&0\\
-\ap\vm&z\id+\ap\am-\idb\vp^t\vm^t\idb&\ap&-\idb\vp^t\\
-\vm&\am&\id&0\\
0&\vm^t\idb&0&1\\
      \end{BMAT}
\right)\,,
\end{equation}
where the blocks on the diagonal are of the size $1$, $r-1$, $r-1$ and $1$ respectively. Here we defined the vectors $\vm$ and $\vp$ containing the oscillators $[\oa_i,\oad_j]=\delta_{ij}$ where $i,j=1,2,\ldots,r-1$ as   
\begin{equation}
  \vm=\left(\begin{array}{cccc}
      \oa_1&\oa_2&\cdots&\oa_{r-1}
     \end{array}
     \right)^t
     \,,\qquad
 \vp=\left(\begin{array}{cccc}
      \oad_1&\oad_2&\cdots&\oad_{r-1}
     \end{array}
     \right)\,.
\end{equation} 
The block matrices $\ap$ and $\am$ are defined as before in \eqref{eq:ApAm} but here they stand for the corresponding matrices of the size $(r-1)\times(r-1)$. Explicitly we have
\begin{equation}\label{eq:App}
\ap=\left(\begin{array}{cccc}
               \oad_{-r+1,1}&\cdots&\oad_{-r+1,r-2}&0\\
               \vdots  &\iddots &0&-\oad_{-r+1,r-2}\\
             \oad_{-2,1}&0&\iddots&\vdots\\
               0&-\oad_{-2,1}&\cdots&-\oad_{-r+1,1}
              \end{array}\right)\,,
\end{equation} 
and
\begin{equation}\label{eq:Amm}
              \am=\left(\begin{array}{cccc}
              - \oa_{1,-r+1}&\cdots&-\oa_{1,-2}&0\\
              \vdots&\iddots &0&\oa_{1,-2}\\
               -\oa_{r-2,-r+1}&0&\iddots&\vdots\\
               0&\oa_{r-2,-r+1}&\cdots&\oa_{1,-r+1}
              \end{array}\right).
    \end{equation} 
In addition to the Lax matrix \eqref{eq:lindec} we introduce the Lax matrix 
\begin{equation}\label{eq:lindec2}
\Lspi_{(-1,\ldots,-1,+1)}(z)=
 \left(\begin{BMAT}[5pt]{c|c:c|c}{c|c:c|c}
z-\vp\vm&\vp\idb&-\vp\idb\am^t&0\\
-\idb \vm&\id&- \am^t&0\\
\ap^t\idb\vm&-\ap^t&z\id+\ap^t\am^t-\idb \vp^t\vm^t\idb&-\vp^t\\
0&0&\vm^t&1\\
      \end{BMAT}
\right)\,,
\end{equation}
which can be obtained from \eqref{eq:lindec} by applying the similarity transformation $B(-1,\ldots,-1,1)$,
cf.~\eqref{eq:invtrans}. We proceed as in Section~\ref{sec:spinlax} and introduce two sets of oscillators to multiply the Lax matrices \eqref{eq:lindec} and \eqref{eq:lindec2} in the matrix space while taking the tensor product in the oscillator space. One finds that the product can be written as
\begin{equation}
 \Lspi^{[1]}(z+s)\Lspi_{(-1,\ldots,-1,+1)}^{[2]}(z-s-\kappa+1)=\mathcal{S}L_s(z)\Gop \mathcal{S}^{-1}\,.
\end{equation} 
Here not all oscillators $[2]$ can be absorbed into the matrix $\Gop$. The latter takes the form
\begin{equation}
\Gop =
 \left(\begin{BMAT}[5pt]{c|c:c|c}{c|c:c|c}
1&0&0&0\\
0&\id&-\am_{[2]}^t&0\\
0&0&\id&0\\
0&0&0&1\\
      \end{BMAT}
\right)\,,
\end{equation} 
which again satisfies $\Gop\Gop'=\ID$ and thus $\Gop$ is a solution to the RTT-relation. It contains $\frac{(r-1)(r-2)}{2}$ annihilation oscillators. Their conjugates have been absorbed by the similarity transformation in the oscillator space 
\begin{equation}
 \mathcal{S}=\exp\left[-\tr(\ap_{[1]}\ap^t_{[2]})-\vp_{[2]}\idb\ap_{[1]}\vm_{[1]}\right]\,.
\end{equation} 
It transforms the oscillators in $\am$  as follows
\begin{equation}
 \mathcal{S}\am_{[2]}^t\mathcal{S}^{-1}=\am_{[2]}^t-\ap_{[1]}\,,\qquad  \mathcal{S}\am_{[1]}\mathcal{S}^{-1}=\am_{[1]}-\ap_{[2]}^t-\vm_{[1]}\vp_{[2]}\idb+\vp^t_{[2]}\vm_{[1]}^t\idb
\end{equation} 
\begin{equation}
 \mathcal{S}\am_{[1]}\mathcal{S}^{-1}=\am_{[1]}-\ap_{[2]}^t-\vm_{[1]}\vp_{[2]}\idb+\vp^t_{[2]}\vm_{[1]}^t\idb
\end{equation} 
while the remaining oscillators transform as
\begin{equation}
 \mathcal{S}\vp_{[1]}\mathcal{S}^{-1}=\vp_{[1]}-\vp_{[2]}\idb \ap_{[1]}\,,\qquad  \mathcal{S}\vm_{[2]}\mathcal{S}^{-1}=\vm_{[2]}+\idb \ap_{[1]}\vm_{[1]}\,,
\end{equation} 
from which follows that 
\begin{equation}
 \mathcal{S}\vp_{[1]}^t\mathcal{S}^{-1}=\vp_{[1]}^t+\idb\ap_{[1]}\vp_{[2]}^t\,,\qquad  \mathcal{S}\vm_{[2]}^t\mathcal{S}^{-1}=\vm_{[2]}^t-\vm_{[1]}^t\idb \ap_{[1]}\,.
\end{equation} 
The final Lax matrix $L_s(x)$ then depends on $\frac{(r+2)(r-1)}{2}$ pairs of oscillators. It can be written as
\begin{equation}\label{eq:quadlaxl}
{\small 
L_s(z)=
 \left(\begin{BMAT}[5pt]{c|c|c}{c|c|c}
(z+s)(z-s-\kappa+1)-\wp\mathcal{L}_s(z)\wm+\frac{1}{4}\wp\idb\wp^t\wm^t\idb\wm&\wp\mathcal{L}_s(z)-\frac{1}{2}\wp\idb\wp^t\wm^t\idb&-\frac{1}{2}\wp\idb\wp^t\\\
-\mathcal{L}_s(z)\wm+\frac{1}{2}\idb\wp^t\wm^t\idb\wm&\mathcal{L}_s(z)-\idb\wp^t\wm^t\idb&-\idb\wp^t\\
-\frac{1}{2}\wm^t\idb\wm&\wm^t\idb&1\\
      \end{BMAT}
\right)\,.}
\end{equation}
where we used the relations $\vp\idb\ap\vp^t=0$ as well as $\vm\idb\ap\vm^t=0$. Here $\mathcal{L}_s(x)$ denotes the spinorial Lax matrix in \eqref{eq:laxspin} of size $2(r-1)\times2(r-1)$ containing $\frac{(r-1)(r-2)}{2}$ pairs of oscillators.
The Lax matrix $L_s$ can be written in the factorised form
\begin{equation}\label{eq:quadl}
{\small
L_s(z)=
 \left(\begin{BMAT}[5pt]{c|c|c}{c|c|c}
1&\wp&-\frac{1}{2}\wp\idb\wp^t\\\
0&\id&-\idb \wp^t\\
0&0&1\\
      \end{BMAT}
\right)
\left(
\begin{BMAT}[5pt]{c|c|c}{c|c|c}
(z+s)(z-s-\kappa+1)  & 0 &  0 \\
0 &\mathcal{L}_s(z) & 0 \\
0& 0 &   1
% \addpath{(1,1,0)ruld}
\end{BMAT}
\right) \left(\begin{BMAT}[5pt]{c|c|c}{c|c|c}
1&0&0\\\
-\wm&\id&0\\
-\frac{1}{2}\wm^t\idb\wm&\wm^t\idb&1\\
      \end{BMAT}
\right)\,,}
\end{equation} 
where
\begin{equation}
 \wp=
\left(
\begin{BMAT}[5pt]{c:c}{c}
\vp_{[2]}\idb  & \vp_{[1]}
% \addpath{(1,1,0)ruld}
\end{BMAT}
\right) \,,\qquad  \wm^t\idb=
\left(
\begin{BMAT}[5pt]{c:c}{c}
\vm_{[1]}^t\idb& \vm_{[2]}^t
% \addpath{(1,1,0)ruld}
\end{BMAT}
\right) \,,
\end{equation} 
 and 
 \begin{equation}
\idb \wp^t=
\left(
\begin{BMAT}[5pt]{c}{c:c}
\idb  \vp_{[1]}^t\\
\vp_{[2]}^t
% \addpath{(1,1,0)ruld}
\end{BMAT}
\right) \,,\qquad  \wm=
\left(
\begin{BMAT}[5pt]{c}{c:c}
\idb\vm_{[2]}\\
\vm_{[1]}
\end{BMAT}
\right) \,.
\end{equation} 

To conclude this section, we remark that the Lax matrix $\mathcal{L}_s$ within $L_s$ is build from the oscillators in \eqref{eq:App} and \eqref{eq:Amm} with subindex $[1]$ which commute with $\wm$ and $\wp$. As such they form a closed $\mathfrak{so}(2(r-1))$ algebra in the representation labelled by $s$.  For $s=0$ and highest weight state  $|0\rangle$, the Fock vacuum of the  annihilation operators $\oa_{i,j}$ in $\mathcal{L}_s$ satisfying $\oa_{i,j}|0\rangle=0$, we find that
\begin{equation}\label{eq:fvac}
 \mathcal{L}_0|0\rangle =z\ID|0\rangle\,.
\end{equation} 
This corresponds to the trivial one-dimensional representation of the Lax matrix in \eqref{eq:slax}. 
The observation \eqref{eq:fvac} allows us to identify $\langle 0|L_0(z)|0\rangle=L(z)$, cf.~\eqref{eq:quadlax}.
As such $L(z)$ is interpreted as a reduction of $L_s$ where the $\mathfrak{so}(2(r-1)$ subalgebra mentioned above is in the trivial representation. Since the oscillators $\wm$ and $\wp$ commute with $\mathcal{L}_s$, this can be seen as a proof that $L(z)$ is a solution to the RTT-relation.
This is similar to the case of $A$-type studied in \cite{Bazhanov:2010jq}.

\subsection{First fundamental representation}\label{sec:fundlax}
In the followig we derive the quadratic solution $ \mathfrak{L}_{n,s}(z)$ corresponding to the representation  $f=(s,s,\ldots,s,n)$ of $\mathfrak{so}(2r)$. In particular for $s=0$ we recover the Lax matrix for the first  fundamental representations $f=(0,\ldots,0,n)$  \cite{Reshetikhin:1986md,Reshetikhin:1986vd} realised in terms of Holstein-Primakoff oscillators. 
To derive the more general solution  $ \mathfrak{L}_{n,s}(z)$ we multiply the Lax matrix \eqref{eq:quadlaxl} containing $\mathcal{L}_s$ with a Lax matrix of the type  \eqref{eq:quadlax} where $s=0$.
For this  we define
\begin{equation}\label{eq:quadlax2}
 L_{(+1,\ldots,+1,-1)}(z)=
 \left(\begin{BMAT}[5pt]{c|c|c}{c|c|c}
1&\wm^t\idb&-\frac{1}{2}\wm^t\idb\wm\\
-\idb\wp^t&z\id-\idb\wp^t\wm^t\idb&-z\wm+\frac{1}{2}\idb\wp^t\wm^t\idb\wm\\
-\frac{1}{2}\wp\idb\wp^t&z\wp-\frac{1}{2}\wp\idb\wp^t\wm^t\idb&z^2+z(2-r-\wp\wm)+\frac{1}{4}\wp\idb\wp^t\wm^t\idb\wm\\
      \end{BMAT}
\right)\,,
\end{equation}
which is obtained from \eqref{eq:quadlax} by acting with the similarity transformation $B(1,\ldots,1,-1)$ in \eqref{eq:invtrans}. Here $\wm$ and $\wp$ denote the vectors defined in \eqref{eq:vecs}.
Multiplying the two solutions \eqref{eq:quadlaxl} and \eqref{eq:quadlax2} we find
\begin{equation}
 L^{[1]}_s(z-x_1) L_{(+1,\ldots,+1,-1)}^{[2]}(z-x_2)=\mathcal{S}\mathfrak{L}_{n,s}^{[1]}(z)\Gop _{[2]}\mathcal{S}^{-1}\,.
\end{equation} 
Here the parameters $x_{1,2}$ are fixed to be
\begin{equation}
 x_1=\frac{2-r-n}{2}\,,\qquad x_{2}=\frac{r+n}{2}\,,
\end{equation} 
% where the role of the parameter $n$ will become clear later in this section. T
while the matrix $\Gop$ reads
\begin{equation}
 \Gop _{[2]}=
 \left(\begin{BMAT}[5pt]{c|c|c}{c|c|c}
1&\wm_{[2]}^t\idb&-\frac{1}{2}\wm^t_{[2]}\idb\wm_{[2]}\\\
0&\id&-\wm_{[2]}\\
0&0&1\\
      \end{BMAT}
\right)\,.
\end{equation} 
Again we determine $\Gop'$ and verify that $\Gop\Gop'=\ID$.
The similarity transformation is defined via 
\begin{equation}
 \mathcal{S}=\exp\left[-\wp_{[1]}\idb\wp_{[2]}^t\right]\,.
\end{equation}
It acts on the oscillators in $(\wm,\wp)$ as 
\begin{equation}
 \mathcal{S}\wm_{[2]}\mathcal{S}^{-1}=\wm_{[2]}+\idb\wp_{[1]}^t\,,\qquad 
 \mathcal{S}\wm_{[2]}^t\mathcal{S}^{-1}=\wm_{[2]}^t+\wp_{[1]}\idb\,,
\end{equation} 
and 
\begin{equation}
 \mathcal{S}\wm_{[1]}\mathcal{S}^{-1}=\wm_{[1]}+\idb\wp_{[2]}^t\,,\qquad 
 \mathcal{S}\wm_{[1]}^t\mathcal{S}^{-1}=\wm_{[1]}^t+\wp_{[2]}\idb\,.
\end{equation} 
The final Lax matrix takes a rather lengthy form and here we only present it in the more concise factorised form
\begin{equation}\label{eq:funfac}
\mathfrak{L}_{n,s}^{[1]}(z)=
 \left(\begin{BMAT}[5pt]{c|c|c}{c|c|c}
1&\wp_{[1]}&-\frac{1}{2}\wp_{[1]}\idb\wp^t_{[1]}\\\
0&\id&-\idb \wp_{[1]}^t\\
0&0&1\\
      \end{BMAT}
\right)\mathrm{D}_s^{[1]}(z)
 \left(\begin{BMAT}[5pt]{c|c|c}{c|c|c}
1&-\wp_{[1]}&-\frac{1}{2}\wp_{[1]}\idb\wp_{[1]}^t\\\
0&\id&\idb \wp_{[1]}^t\\
0&0&1\\
      \end{BMAT}
\right)\,,
\end{equation} 
where the middle part is written as
\begin{equation}
 \mathrm{D}_s^{[1]}(z)= \left(\begin{BMAT}[5pt]{c|c|c}{c|c|c}
(z-x_1)(z-x_1-r+2)&0&0\\\
-\mathcal{L}_s(z-x_1)\wm_{[1]}&(z-x_2)\mathcal{L}_s(z-x_1)&0\\
-\frac{1}{2}\wm^t_{[1]}\idb\wm_{[1]}&\wm^t_{[1]}\idb(z-x_2)&(z-x_2)(z-x_2-r+2)\\
      \end{BMAT}
\right)\,.
\end{equation} 
Here $\mathcal{L}_s$ denotes the spinorial Lax matrix in \eqref{eq:laxspin}. With the same argument as outlined in the previous subsection we can consider the trivial representation with $s=0$. Then the middle part of the Lax matrix in \eqref{eq:funfac} simplifies and yields
\begin{equation}
 \mathrm{D}_0^{[1]}(z)= \left(\begin{BMAT}[5pt]{c|c|c}{c|c|c}
(z-x_1)(z-x_1-r+2)&0&0\\\
-\wm_{[1]}(z-x_1)&\id(z-x_1)(z-x_2)&0\\
-\frac{1}{2}\wm^t_{[1]}\idb\wm_{[1]}&\wm^t_{[1]}\idb(z-x_2)&(z-x_2)(z-x_2-r+2)\\
      \end{BMAT}
\right)\,.
\end{equation} 
The matrix $\mathfrak{L}_{n,s}$ contains $\frac{(r-1)(r+2)}{2}$ pairs of oscillators while after taking $s=0$ we reduce the number of oscillators by $\frac{(r-1)(r-2)}{2}$ and effectively remain with $2(r-1)$ pairs of oscillators in $\mathfrak{L}_{n}$. By construction these are solutions to the RTT-relation.

In the latter case with $s=0$ we can compare our result to the quadratic solutions known in the literature corresponding for the symmetric generalisations of the first fundamental representation \cite{Reshetikhin:1986md} where the generators $M_{AB}$, as discussed in Appendix~\ref{app:gens}, were used. In our basis it can be written in terms of the $\mathfrak{so}(2r)$ generators $F_{ab}$ as
\begin{equation}\label{eq:fundlax}
 \mathfrak{L}_{n}(z)=z^2+zE_{ab}F_{ba}+E_{ab}G_{ba}\,,
\end{equation}  
where $G_{ab}$ can be expressed in terms of the generators $F_{ab}$ as 
\begin{equation}\label{eq:GG}
G_{ab}=\frac{1}{2}F_{cb}F_{ac}+\frac{\kappa}{2}F_{ab}-\frac{1}{4}\left((\kappa-1)^2+2\kappa n+n^2\right)\delta_{ab}\,.
\end{equation} 
The constraint satisfied by the generators is cubic and reads
\begin{equation}\label{eq:charf}
 \left(F_{ab}-\delta_{ab}\right) \left(F_{bc}-n\delta_{bc}\right) \left(F_{cd}+(n-2\kappa)\delta_{cd}\right)=0\,,
\end{equation} 
cf.~\cite{Reshetikhin:1986md}.
We have verified \eqref{eq:GG} and \eqref{eq:charf} for $r=3,4,5$ using a computer algebra program. We refer the reader to \cite{Karakhanyan:2017wyl} where the constraints on  the Lax matrix are studied in general.

The resulting Holstein-Primakoff realisation is labelled by the weights $f=(0,\ldots,0,n)$ with the Fock vacuum as the highest weight state. It is finite dimensional for $n\in \mathbb{N}_0$. Inserting the first  fundamental representation in terms of matrices $F_{ab}=E_{ab}-E_{-b,-a}$ into \eqref{eq:fundlax} we recover $R(z-\frac{\kappa}{2})$.
We further note that we recover the degenerate Lax matrix \eqref{eq:quadlax} by taking the limit
\begin{equation}
 L(z)=\lim_{n\to\infty}\frac{\mathfrak{L}_n(z+x_1)}{n}\left(\begin{BMAT}[5pt]{c|c|c}{c|c|c}
n&0&0\\\
0&-\id&0\\
0&0&n^{-1}\\
      \end{BMAT}
\right)\,.
\end{equation} 

To end this section we remark that one can show that the linear order of $\mathfrak{L}_{n,s}$ yields a representation of $\mathfrak{so}(2r)$ in Holstein-Primakoff form with the Fock vacuum as a highest weight state and the representation labels $f=(s,\ldots,n)$.  For the equivalent Lax matrix in Jordan-Schwinger form we refer the reader to \cite{Isaev:2015hak}.
We expect that the Lax matrix corresponding to the representation labeled via $f=(-s,s,\ldots,s,n)$ can be obtained in the same way by replacing the Lax matrix $\mathcal{L}_s$ with the one for the other spinor note as discussed in Section~\ref{sec:spinlax}. 

\section{Towards Q-operators for $\mathfrak{so}(2r)$ invariant spin chains}\label{sec:Qops}
This section is devoted to the definition of transfer matrices from monodromies containing the solutions we have introduced in the previous section. 
The fundamental transfer matrix is well known. It can explicitly be defined as
\begin{equation}\label{eq:transm}
 \mathbf{T}(x)=\tr_a \mathbf{D}_a \mathbf{R}_{a1}(x)\mathbf{R}_{a2}(x)\cdots \mathbf{R}_{aN}(x)\,,
\end{equation} 
where we introduced the diagonal twist matrix only acting non-trivially in the auxiliary space 
\begin{equation}\label{eq:twist}
 \mathbf{D}=\diag\left(\exp[-\phi_r],\ldots,\exp[-\phi_1],\exp[+\phi_1],\ldots,\exp[+\phi_r]\right)\,.
\end{equation} 
Here the twist parameters are complex variables $\phi_i\in \mathbb{C}$ with $i=1,\ldots,r$ and the twist matrix enjoys the property $\mathbf{D}\mathbf{D}'=\ID$. As a consequence of the Yang-Baxter equation, the transfer matrix constructed in this way commutes with itself at different values of the spectral parameter
\begin{equation}
 [\mathbf{T}(x),\mathbf{T}(y)]=0\,.
\end{equation} 
Below we construct further members of the commuting family of operators from the minimal oscillator type solutions introduced in Section~\ref{sec:minimal}. Here the oscillator space will take the role of the auxiliary space. We expect that these transfer matrices  are the Q-operators corresponding to the nodes of the Dynkin diagram such that their eigenvalues are polynomials with zero's given by the Bethe roots of the appropriate nesting level.

\subsection{Spinor type}
We can construct a Q-operator from the spinor type solution in \eqref{eq:laxlin} that commutes with the fundamental transfer matrix \eqref{eq:transm} by multiplying $N$ Lax matrices $\Lspi(x)$ in the oscillator space while taking the tensor product in the matrix space. The Q-operator is then given as the regularised trace over the oscillator space. It reads
\begin{equation}
 \mathbf{Q}_s(x)= \tr_{\mathrm{osc}} D_s\, \Lspi(x)\otimes\Lspi(x)\otimes\ldots\otimes\Lspi(x)
\end{equation} 
where we introduced the twist (regulator) 
\begin{equation}
 D_s=\exp\left[\sum_{1\leq i<j\leq r}(\phi_i+\phi_j)\oad_{-j,i}\oa_{i,-j}\right]\,.
\end{equation} 
We can verify that the adjoint action of the twist matrix $\mathbf{D}$ on the Lax matrix  $\Lspi(x)$ can be absorbed by the twist in the oscillator space
% \begin{equation}\label{eq:com}
%   \mathbf{D}\Lspi(z)\mathbf{D}^{-1}= D_s^{-1}\Lspi(z)D_s\,.
% \end{equation} 
\begin{equation}\label{eq:com}
 \left[\Lspi(z),\mathbf{D}\otimes D_s\right]=0
\end{equation} 
As a consequence of the RTT-relation and the property \eqref{eq:com} it follows that the operator $\mathbf{Q}_s(x)$ 
 belongs to the commuting family of operators.
In total we can define $2^r$ Q-operators using the transformation given in \eqref{eq:invtrans}. They are labeled by the vector $\vec\alpha$ and defined via
\begin{equation}
 \mathbf{Q}_{s}(x;\vec\alpha)= \left(B(\vec\alpha)\otimes\ldots\otimes B(\vec\alpha)\right) \mathbf{Q}_s(x)\left(B(\vec\alpha)\otimes\ldots\otimes B(\vec\alpha)\right)|_{\phi_i\to \alpha_i\phi_i}\,.
\end{equation} 
All of them commute with the transfer matrix $\mathbf{T}(x)$ at different points of the spectral parameter. This can be shown using the relation
\begin{equation}
 \left(B(\vec\alpha)\otimes\ldots\otimes B(\vec\alpha)\right)\mathbf{T}(x)\left(B(\vec\alpha)\otimes\ldots\otimes B(\vec\alpha)\right)=\mathbf{T}(x)|_{\phi_i\to \alpha_i\phi_i}\,,
\end{equation} 
which follows from the invariance of the R-matrix and the explicit form of the twist in \eqref{eq:twist}.
\subsection{First fundamental type}
A similar construction can be done for the oscillator type solutions on the first fundamental node.
Here the Q-operator is defined via
\begin{equation}\label{eq:Qf}
 \mathbf{Q}_f(x)=\tr_{\mathrm{osc}} D_f\, L(x)\otimes L(x)\otimes\ldots\otimes L(x)\,.
\end{equation} 
The Lax matrix $L(x)$ was introduced in \eqref{eq:quadlax} and the twist (regulator) reads
\begin{equation}
 D_f=\exp\left[\sum_{i=1}^{r-1}(\phi_r+\phi_i)\oad_{i}\oa_{i}\right]\exp\left[\sum_{i=1}^{r-1}(\phi_r-\phi_i)\oad_{-i}\oa_{-i}\right]\,.
\end{equation} 
Again one can verify that it obeys the corresponding relation as presented for the spinorial case in \eqref{eq:com} and conclude that it commutes with the transfer matrix $\mathbf{T}(x)$.
Further Q-operators on the same node can be obtained from the one defined in \eqref{eq:Qf} using the similarity transformations introduced in Section~\ref{sec:fundR}. 
Noting that  $B(-1,\ldots,-1)=\idb$, see \eqref{eq:invtrans}, we define the operator
\begin{equation}
 \bar{\mathbf{Q}}_f(x)=\left(\idb\otimes\ldots\otimes\idb\right) \mathbf{Q}_f(x)\left(\idb\otimes\ldots\otimes\idb\right)|_{\phi_i\to -\phi_i}\,.
\end{equation} 
Further using the transformation in \eqref{eq:invtrans2} we define
\begin{equation}
 {\mathbf{Q}}_f(x;i)=\left(\tilde B_{ir}\otimes\ldots\otimes \tilde B_{ir}\right) \mathbf{Q}_f(x)\left(\tilde B_{ir}\otimes\ldots\otimes \tilde B_{ir}\right)|_{\phi_r\leftrightarrow \phi_i}\,,
\end{equation} 
and 
\begin{equation}
 \bar{\mathbf{Q}}_f(x;i)=\left(\tilde B_{ir}\otimes\ldots\otimes \tilde B_{ir}\right) \bar{\mathbf{Q}}_f(x)\left(\tilde B_{ir}\otimes\ldots\otimes \tilde B_{ir}\right)|_{\phi_r\leftrightarrow \phi_i}\,,
\end{equation} 
with $1\leq i\leq r-1$. Thus we found in total $2r$ Q-operators for the first fundamental node.
It can be shown using a similar argument as done for the spinor case that the Q-operators obtained from $\mathbf{Q}_f$  commute with the transfer matrix. 
We expect that the construction of the transfer matrices for the remaining Lax matrices can be done analogously. 

\subsection{Some QQ-relations for $D_4$} In this subsection we present some QQ-relations which support the identifications of the operators introduced in this section with Baxter Q-operators. We focus on the first non-trivial case, i.e. $r=4$. The quantum space for a given length $N$ of the spin chain is of the size $8^N$ which complicates numeric checks even at low length.
For $N=1,2,3$ we were able to check the following QQ-relations:\\[0.2cm]
$\bullet$ $1$st spinor node:
\begin{equation}
 e^{\phi_1}\mathbf{Q}_s^{[x-1]}(z;\{2\})\mathbf{Q}_s^{[x]}(z;\{1\})-e^{\phi_2}\mathbf{Q}_s^{[x]}(z;\{2\})\mathbf{Q}_s^{[x-1]}(z;\{1\})=\left(e^{\phi_1}-e^{\phi_2}\right)\mathbf{A}^{[x]}(z)
\end{equation} 
$\bullet$ $2$nd spinor node:

\begin{equation}
 e^{\phi_1}\mathbf{Q}_s^{[x-1]}(z;\emptyset)\mathbf{Q}_s^{[x]}(z;\{1,2\})-e^{-\phi_2}\mathbf{Q}_s^{[x]}(z;\emptyset)\mathbf{Q}_s^{[x-1]}(z;\{1,2\})=\left(e^{\phi_1}-e^{-\phi_2}\right)\mathbf{A}^{[x]}(z)
\end{equation} 
$\bullet$ $1$st fundamental node:

\begin{equation}
 e^{\phi_3}\mathbf{Q}_f^{[x+1]}(z)\mathbf{Q}_f^{[x]}(z;3)-e^{\phi_4}\mathbf{Q}_f^{[x]}(z)\mathbf{Q}_f^{[x+1]}(z;3)=\left(e^{\phi_3}-e^{\phi_4}\right)\mathbf{A}^{[x]}(z)\mathbf{Q}_0^{[x]}(z)
\end{equation} 
for any constant $x$. Here we employed the standard notation $f^{[x]}(z)=f(z+x)$ and denoted the Q-operators via $\mathbf{Q}_s(z;I)=\mathbf{Q}_s(z;\vec\alpha)$ where $I=\{i_1,\ldots i_k\}$ denotes the positions of $k$ minus signs in $\vec\alpha$, e.g. $\{2\}\leftrightarrow (1,-1,1,\ldots,1)$. Further $\mathbf{Q}_0(z)$ denotes the diagonal Q-operator
\begin{equation}
 \mathbf{Q}_0(z)=(z+1)^N\,.
\end{equation} 
It may be viewed as the transfer matrix built from the Lax matrix \eqref{eq:laxspin} when taking the trivial representation in the auxiliary oscillator space.
We expect that the matrix $\mathbf{A}(z)$ is a Q-operator that corresponds to the middle node on the Dynkin diagram of $D_4$, cf.~Figure~\ref{fig:dr}, for which we do not have a Lax matrix construction.  We have also checked that the operators commute among themselves at different values of the spectral parameter for the cases considered.
Further functional relations can be obtained by applying the similarity transformations \eqref{eq:invtrans} and \eqref{eq:invtrans2} in the quantum space and subsequently interchanging the twists as discussed previously. Similar relations were proposed in \cite{Masoero:2015lga} in connection to the 
ODE/IM correspondence \cite{Dorey:2007zx}. 
We plan to study the functional relations for $D$-type spin chains in further detail in an upcoming publication \cite{toappear}.

\section{Conclusion}\label{sec:conc}
In this article we obtained four families of solutions to the Yang-Baxter equation of $\mathfrak{so}(2r)$. Two of them seem to be completely new and important for the construction of Q-operators for $\mathfrak{so}(2r)$ spin chains. Two others were known but the realisation of the $\mathfrak{so}(2r)$ algebra in terms of the oscillators has not been considered in the literature. They can be used to construct transfer matrices.
We found that it is natural to work in the basis considered in \cite{Arnaudon2006} where the degenerate Lax matrices have the spectral parameter at  leading order on the diagonal and the Yangian commutation relations can be written in a rather compact form.
Further, we showed several factorisation formulas among the Lax matrices. All of them are resembling the form of the factorisation that appeared in \cite{Bazhanov:1998dq} and later in the construction of Q-operators for  spin chains in \cite{Bazhanov:2010jq,Bazhanov:2010ts}. Also we have shown that vice versa they are related by taking a limit as familiar for $\mathfrak{sl}(2)$, see e.g. \cite{sklyanin00,Bazhanov:2010ts}, or \cite{Bazhanov:2008yc} for the trigonometric case of  $\mathfrak{sl}(2|1)$. We expect that the transfer matrices defined in Section~\ref{sec:Qops} are the Q-operators corresponding to the extremal nodes of the Dynkin diagram. This is supported by our case study of the QQ-relations for $r=4$ and low lengths.
To proof directly that the operators constructed here are indeed the Q-operators one may either show that they obey TQ-relations, cf.~\cite{Frenkel:2013uda}, or diagonalise them directly using the algebraic Bethe ansatz, c.f.~\cite{Bazhanov:2010jq,Bazhanov:2010ts} and \cite{Frassek:2015qra} for the case of $A$-type Lie algebras. Further 
it would be interesting to compare our results to the  construction in \cite{Frenkel:2013uda} using the prefundamental representation \cite{Hernandez:2011ama}. See also \cite{Boos:2016,Boos:2017mqq} where the relation between oscillator and prefundamental representations is discussed for the trigonometric setup with $A$-type. For the corresponding Lax matrices including the supersymmetric extension we refer the reader to \cite{Tsuboi:2019vvv} and references therein.

There are several directions that we plan to explore in the future.
In particular the construction of all Q-operators and T-operators using the oscillators as introduced here and a full derivation of the functional relations among them is an interesting goal to pursue. For this purpose, it may be advantageous to also study the relation to the construction using characters as done in \cite{Kazakov:2010iu} for $A$-type. The factorisation formulas found here are expected to yield proofs of certain functional relations of Q-operators with transfer matrices. 
We expect that our construction can be generalised to the case where the quantum space (matrix space) is in spinor representation as well as to symmetric analogs of the fundamental representation, see \cite{Frassek:2011aa} for the case of $A_r$. This is also suggested by the factorised form of the Lax matrices. 
It seems however difficult to obtain solutions corresponding to non-extremal nodes of the Dynkin diagram. So far our attempts of deriving them via fusion have failed. To understand such representations but also to achieve a more complete classification  we plan to evaluate the Lax matrices from the shifted Yangian \cite{Braverman:2016pwk} as done for $A$-type in \cite{FPT}. We plan to report on this subject in an upcoming publication. It would further be interesting to verify  the spectral properties of the Lax matrices as predicted in \cite{Nekrasov:2012xe,Nekrasov:2013xda}. This has been studied in \cite{Frassek:2018try} for  case of $A$-type Lie algebras.
Other directions include the study of oscillator type K-matrices as done in \cite{Frassek:2015mra,Baseilhac:2017hoz,Tsuboi:2018gfd}, the generalisation of the obtained solutions to the trigonometric case and of course the study of oscillator type Lax matrices, Q-operators and functional relations that emerge from  Yangians corresponding to other Lie algebras.

\section*{Acknowledgements}
I like to thank Stefano Negro, Alexander Tsymbaliuk and especially Vasily Pestun for useful comments and suggestions as well as Alexander Tsymbaliuk for comments on the manuscript. I also thank the anonymous referees for their comments. I am thankful to the support of the  visitor program of the IH\'ES where large parts of this work were carried out.
I also acknowledge the support of the DFG Research Fellowships Programme 416527151 and thank the organisers of \href{https://lapth.cnrs.fr/conferences/RAQIS/RAQIS18/}{RAQIS'18} where parts of this work were presented. 

% \newpage
\appendix
\section{Generators and commutation relations}\label{app:gens}
It is straightforward to compute the action of the similarity transformation $\Sop$ in \eqref{eq:simtrans} on the matrix elements $e_{AB}$. We find
\begin{align}
 \Sop e_{i,j}\Sop^{-1}&=\frac{1}{2}\left(E_{-i,-j}-E_{-i,j}-E_{i,-j}+E_{i,j}
\right)\,,\\  \Sop e_{i,r+j}\Sop^{-1}&=\frac{i}{2}\left(-E_{-i,-j}-E_{-i,j}+E_{i,-j}+E_{i,j}\right)
 \\
 \Sop e_{r+i,j}\Sop^{-1}&=\frac{i}{2}\left(E_{-i,-j}-E_{-i,j}+E_{i,-j}-E_{i,j}
\right)\,,\\ \Sop e_{r+i,r+j}\Sop^{-1}&=\frac{1}{2}\left(E_{-i,-j}+E_{-i,j}+E_{i,-j}+E_{i,j}
\right)\,,
\end{align}
where $i,j=1,\ldots,r$. It now follows immediately that 
\begin{equation}
\begin{split}
 \sum_{k_1,k_2=0}^1\left(\Sop e_{rk_1+i,rk_2+j}\Sop^{-1}\right)\otimes  \left(\Sop e_{rk_2+j,rk_1+i}\Sop^{-1}\right)=\sum_{k_1,k_2\in\{+1,-1\}}
 E_{k_1i,k_2j}\otimes E_{k_2j,k_1i}\,,
 \end{split}
\end{equation} 
cf.~Section~\ref{sec:fundR},  and we find $\left(\Sop\otimes\Sop\right) \Kop \left(\Sop^{-1}\otimes\Sop^{-1}\right) =\Qop$ after summing over the indices $i,j$.

The Lax matrices for $\mathfrak{so}(2r)$ are commonly written in terms of the generators
\begin{equation}
 [M_{AB},M_{CD}]=\delta_{AD}M_{BC}+\delta_{BC}M_{AD}-\delta_{AC}M_{BD}-\delta_{BD}M_{AC}\,,
\end{equation} 
with $M_{AB}=-M_{BA}$. 
In Section~\ref{sec:factorisation} we defined the generators $F_{ab}$ which are related to $M_{AB}$ via
\begin{equation}
\sum_{a,b}E_{ab}F_{ba}=\sum_{A,B}
 \Sop e_{AB}\Sop^{-1}M_{BA}\,.
\end{equation} 
This identification can be written in terms of components as
\begin{align}
  F_{-i,-j}&=\frac{1}{2}\left(M_{i,j}+i M_{i,j+r}-i M_{i+r,j}+M_{i+r,j+r}\right)\,,
  \\
 F_{-i,j}&=\frac{1}{2}\left(-M_{i,j}+i M_{i,j+r}+i M_{i+r,j}+M_{i+r,j+r}\right)\,,\\
 F_{i,-j}&=\frac{1}{2}\left(-M_{i,j}-i M_{i,j+r}-i M_{i+r,j}+M_{i+r,j+r}\right)\,,\\
 F_{i,j}&=\frac{1}{2}\left(M_{i,j}-i M_{i,j+r}+i M_{i+r,j}+M_{i+r,j+r}\right)\,.
\end{align}

{
% \small
\bibliographystyle{utphys2}
\bibliography{refs}
}
% rouven.frassek@phys.ens.fr

\end{document}